\documentstyle[12pt,fleqn]{article}
\def\beq{\begin{equation}}
\def\eeq{\end{equation}}

\begin{document}

\begin{center}
Physical Review A {\bf54}, 2685 (1996)\\[1cm]
{\Large Collective tests for quantum nonlocality} \\[2cm]
Asher Peres$^*$ \\[7mm]
{\sl Department of Physics, Technion---Israel Institute of
Technology, 32\,000 Haifa, Israel}\\[1cm]
\end{center}\vfill

\noindent{\bf Abstract}\bigskip

Pairs of spin-$1\over2$ particles are prepared in a Werner state
(namely, a mixture of singlet and random components). If the random
component is large enough, the statistical results of spin measurements
that may be performed on each pair separately can be reproduced by an
algorithm involving local ``hidden'' variables. However, if several such
pairs are tested simultaneously, a violation of the
Clauser-Horne-Shimony-Holt inequality may occur, and no local hidden
variable model is compatible with the results.  \vfill

\vfill

\noindent $^*$\,Electronic address: peres@photon.technion.ac.il

\vfill\newpage

\begin{center}{\bf I. INTRODUCTION}\end{center}

From the early days of quantum mechanics, the question has often been
raised whether an underlying ``subquantum'' theory, that would be 
deterministic or even stochastic, was viable. Such a theory would
presumably involve additional ``hidden'' variables, and the statistical
predictions of quantum theory would be reproduced by performing suitable
averages over these hidden variables.

A fundamental theorem was proved by Bell~\cite{Bell}, who showed that if
the constraint of {\it locality\/} was imposed on the hidden variables
(namely, if the hidden variables of two distant quantum systems would
themselves be separable into two distinct subsets), then there was an
upper bound to the correlations of results of measurements that could be
performed on the two distant systems. That upper bound, mathematically
expressed by Bell's inequality~\cite{Bell}, is violated by some states
in quantum mechanics, for example the singlet state of two
\mbox{spin-$1\over2$} particles.

A variant of Bell's inequality, more general and more useful for
experimental tests, was later derived by Clauser, Horne, Shimony, and
Holt (CHSH)~\cite{chsh}. It can be written

\beq |\langle{AB}\rangle+\langle{AB'}\rangle+\langle{A'B}\rangle
  -\langle{A'B'}\rangle|\leq 2. \label{CHSH}\eeq
On the left hand side, $A$ and $A'$ are two operators that can be
measured by the first observer, conventionally called Alice. These
operators do not commute (so that Alice has to choose whether to
measure $A$ or $A'$) and each one is normalized to unit norm (the norm 
of an operator is defined as the largest absolute value of any of its
eigenvalues). Likewise, $B$ and $B'$ are two normalized noncommuting
operators, any one of which can be measured by another, distant
observer (Bob). Note that each one of the {\it expectation\/} values in
Eq.~(\ref{CHSH}) can be calculated by means of quantum theory, if the
quantum state is known, and is also experimentally observable, by
repeating the measurements sufficiently many times, starting each time
with identically prepared pairs of quantum systems.

The validity of the CHSH inequality, for {\it all\/} combinations of
measurements independently performed on both systems, is a necessary 
condition for the possible existence of a local hidden variable (LHV)
model for the results of these measurements. It is not in general a
sufficient condition, except in some simple cases, for example when each
observer is testing a two-state system, and has only two alternative
tests to choose from~\cite{Fine}. For more general situations,
counter\-examples can be found, such that the inequality~(\ref{CHSH})
holds for any two pairs of correlation coefficients, and yet the
nonexistence of a LHV model can be proved~\cite{GM}. The purpose of the
present article is to show that, even if a well defined LHV model
exists, that reproduces all the statistical properties of pairs of
particles when each pair is tested separately, there may be no extension
of such a model that is valid when several pairs are tested
simultaneously.

Note that the difficulty appears only in the case of {\it  mixed\/}
quantum states. For pure states, it is easily shown that the CHSH
inequality is violated by any non-factorable state~\cite{Capasso,GiPe},
while on the other hand  a factorable state trivially admits a
(contextual) LHV model~\cite{Bell66}. For a pair of spin-$1\over2$
particles with a given mixed density matrix, there is an explicit
formula~\cite{H3} which gives the maximum value of the left hand side of
Eq.~(\ref{CHSH}), for any measurements that can be chosen by Alice and
Bob [see Eq.~(\ref{M}) below]. However, even if that maximum value is
less than~2, so that the CHSH inequality holds, this does not prove as
yet that a LHV model is admissible, as will be shown in this article.

For quantum systems whose states lie in higher dimensional vector
spaces, even less is known~\cite{MS}. Some time ago, Werner~\cite{Wer}
constructed a density matrix $\rho_{\rm W}$
%
%
%
for a pair of spin-$j$ particles, with paradoxical properties. Werner's
state $\rho_{\rm W}$ cannot be written as a sum of direct products of
density matrices, $\sum_j c_j\,  \rho_{{\rm A}j}\otimes\rho_{{\rm B}j}$,
where $\rho_{{\rm A}j}$ and $\rho_{{\rm B}j}$ refer to the two distant
particles (the indices A and B stand for Alice and Bob, respectively).
Therefore, genuinely quantal correlations are involved in $\rho_{\rm
W}$. Nevertheless, for any pair of ideal local measurements performed on
the two particles, the correlations derived from $\rho_{\rm W}$ not only
satisfy the CHSH inequality, but, as Werner showed~\cite{Wer}, it is
possible to introduce an explicit LHV model that correctly reproduces
all the observable correlations for these ideal measurements.

For a pair of spin-$1\over2$ particles, Werner's state is

\beq
 \rho_{\rm W}=\mbox{$1\over2$}\,(\rho_{\rm singlet}+\mbox{$1\over4$}\,
 {\leavevmode\hbox{\small1\kern-3.8pt\normalsize1}}),
  \label{Werner}\eeq
namely, an equal weight mixture of a singlet state (which maximally
violates the CHSH inequality) and a totally uncorrelated random state.
Note that this mixture is rotationally invariant. A manifestly
nonclassical property of $\rho_{\rm W}$ was discovered by
Popescu~\cite{Pop94}, who showed that such a particle pair could be used
for teleportation of a quantum state~\cite{telep}, albeit with a
fidelity lesser than if a pure singlet were employed for that purpose.
This nonclassical property came as a surprise, and it was the first
indication that the existence of a formal LHV model was not a complete
description of this system. Indeed, the abstract LHV model that was
proposed by Werner deals only with pairs of ideal measurements of the
von Neumann type. It is not a complete theory, because it does not
predict what happens if other measuring methods are chosen. In
particular, Werner's algorithm becomes ambiguous for spin
$>\,{1\over2}$, when we consider the measurement of projection operators
of rank 2 or higher~\cite{Shimony}. The algorithm must then be
supplemented by further rules.

This ambiguity was exploited by Popescu~\cite{Pop95} in the following
way. Instead of measuring complete sets of orthogonal projection
operators of rank 1, as discussed in Werner's article, Alice and Bob
first measure suitably chosen (and mutually agreed) projection operators
of rank 2, say $P_{\rm A}$ and $P_{\rm B}$. If one of them gets a null
result, the experiment is considered to have failed, and they test
another Werner pair. Only if both Alice and Bob find the result 1 for
$P_{\rm A}$ and $P_{\rm B}$, they proceed by independently choosing
projection operators of rank 1, in the subspaces spanned by $P_{\rm A}$
and $P_{\rm B}$, respectively.  Popescu then shows that if the initial
Hilbert space (for each particle) has dimension 5 or higher, the
correlation of the final results violates the CHSH inequality. In other
words, Werner's hidden variable model, which worked for single ideal
measurements, is incapable of reproducing the results of several {\it
consecutive\/} measurements (and of course no other hidden variable
model would be acceptable).

Popescu's measuring method~\cite{Pop95} does not lead to a violation of
the CHSH inequality in spaces having fewer than $5\times5$ dimensions.
Nonetheless, such a violation can be produced with the simplest Werner
pairs, made of two-state systems, by combining several pairs together.
In order to achieve this result, Alice and Bob must first ``purify'' the
Werner state, and distill, from a large set of Werner pairs, a subset of
almost pure singlets~[15--18]. In the discussion of that purification
procedure, the notion of ``Werner state'' has to be generalized from its
original definition (\ref{Werner}) to

\beq \rho_{\rm W}=x\,\rho_{\rm singlet}+\mbox{$1\over4$}\,(1-x)\,
 {\leavevmode\hbox{\small1\kern-3.8pt\normalsize1}}.\eeq
This state consists of a singlet fraction $x$ and a random (totally
uncorrelated) fraction $(1-x)$. States of this type were first
considered by Blank and Exner \cite{Exner}. Note that the random
fraction $(1-x)$ also includes singlets, mixed in equal proportions with
the three triplet components. Another commonly used measure of
entanglement is the ``fidelity''

\beq F=(3x+1)/4, \eeq
which is the {\it total\/} fraction of singlets [15--18].

In the present work, I shall not consider the fractional distillation of
singlets---a multi\-stage process---but the result of a single
simultaneous observation of several Werner pairs. Namely, if there are
$n$ such pairs, Alice and Bob perform their tests on quantum systems
consisting of $n$ particles (each system is described by a vector space
of dimension $2^n$). To understand why new results may be obtained by
means of such collective tests, let us recall how the statistical
interpretation of the quantum formalism is related to actual statistical
tests. When we say that a physical system has a density matrix $\rho$,
this means that we may mentally construct a {\it Gibbs ensemble\/} of
such systems, namely an infinite set of conceptual replicas of it, all
prepared in the same way \cite{ajp}. This mental process is not the same
thing as actually preparing a large number of such systems, say $N$ of
them. The latter preparation gives a {\it Maxwell ensemble\/} (for
example, a gas made of $N$ identical molecules). If we test individually
the various members of a Maxwell ensemble, we may approach, in the limit
$N\to\infty$, the statistical properties computed for the Gibbs
ensemble. I emphasize that the latter is a pure theoretical construct,
needed for the sole purpose of statistical reasoning.

Now, once there actually is a Maxwell ensemble, we may also test its
constituent systems two by two, or three by three, etc. In that case,
we effectively consider a new kind of physical system, that consists of
two, or three, or more, of the former ``physical systems.'' If the
mathematical representation of the states of the old systems was a
density matrix $\rho$, then the representation of the new systems is
given by a tensor product, such as $\rho\otimes\rho$, or
$\rho\otimes\rho\otimes\rho$, etc. The purpose of this article is to
show that even if the density matrices $\rho$ obey the CHSH inequality,
it is possible that $\rho\otimes\rho$, or $\rho\otimes\rho\otimes\rho$,
etc., violate that inequality, when we measure suitably chosen
operators.\bigskip

\begin{center}{\bf II. PROTOCOL FOR COLLECTIVE TESTS} \end{center}

In the case of Werner pairs that are considered here, each one of the
two observers has $n$ particles (one particle from each Werner pair).
The two observers then proceed as follows. First, they subject their
$n$-particle systems to suitably chosen local unitary transformations,
$U$, for Alice, and $V$, for Bob. (This is always possible, in
principle, by using a multiport~\cite{Reck} or a similar device.) Then,
they test whether each one of the particles labelled 2, 3, \ldots, $n$,
has spin up (for simplicity, it is assumed that all the particles are
distinguishable, and can be labelled unambiguously). Note that any other
test that they can perform is unitarily equivalent to the one for spins
up, as this involves only a redefinition of the matrices $U$ and $V$. If
any one of the $2(n-1)$ particles tested by Alice and Bob shows spin
down, the experiment is considered to have failed, and the two observers
must start again with $n$ new Werner pairs. A similar elimination of
``bad'' samples is also inherent to Popescu's protocol~\cite{Pop95}, or
to any experimental procedure where a failure of one of the detectors to
fire is handled by discarding the results registered by all the other
detectors:  only when {\it all\/} the detectors fire are their results
included in the statistics. This obviously requires an exchange of {\it
classical\/} information between the observers.

Note that, instead of the unitary transformations $U$ and $V$, Alice and
Bob could use more general {\it nonunitary\/} transformations, involving
selective absorption~\cite{Gisin96}. The latter can sometimes be used to
enhance nonlocality, but they would not help in the present case,
because Werner states are rotationally symmetric. Still another
possibility would be for Alice and Bob to use a positive operator valued
measure (POVM)~\cite{qt}, by including in their apparatuses auxiliary
quantum systems, independently prepared by each one of them, and then
performing local unitary transformations and tests on the {\it
combined\/} systems. In the present work, such a strategy was examined,
as a possible alternative to the simpler one discussed above, and it was
found that no advantage resulted from the use of a POVM. This is likely
due to the rotational symmetry of Werner states. I shall therefore
restrict the following discussion to what happens after {\it unitary\/}
transformations $U$ and $V$ have been performed on the $n$ particles
held by each observer.

The calculations shown below will refer to the case $n=3$, for
definiteness.  The generalization to any other value of $n$ is
straightforward. Spinor indices, for a single \mbox{spin-$1\over2$}
particle, will take the values 0 (for the ``up'' component of spin) and
1 (for the ``down'' component). The 16 components of the density matrix
of a Werner pair, consisting of a singlet fraction $x$ and a random
fraction $(1-x)$, are, in the standard direct product basis:

\beq \rho_{mn,st}=x\,S_{mn,st}+(1-x)\,\delta_{ms}\,\delta_{nt}\,/4,\eeq
where the indices $m$ and $s$ refer to Alice's particle, $n$ and $t$
to Bob's particle, and where the density matrix for a pure singlet is
given by

\beq S_{01,01}=S_{10,10}=-S_{01,10}=-S_{10,01}=\mbox{$1\over2$},
\label{singlet}\eeq
and all the other components of $S$ vanish.

When there are three Werner pairs, their combined density matrix is a
direct product $\rho\otimes\rho'\otimes\rho''$, or explicitly,
$\rho_{mn,st}\,\rho_{m'n',s't'}\,\rho_{m''n'',s''t''}$. The result of
the unitary transformations $U$ and $V$ is

\beq \rho\otimes\rho'\otimes\rho''\to
 (U\otimes V)\,(\rho\otimes\rho'\otimes\rho'')\,
 (U^\dagger\otimes V^\dagger). \label{newrho}
\eeq
Explicitly, with all its indices, the $U$ matrix satisfies the unitarity
relation

\beq \sum_{mm'm''}U_{\mu\mu'\mu'',mm'm''}\;
 U^*_{\lambda\lambda'\lambda'',mm'm''}=
 \delta_{\mu\lambda}\;\delta_{\mu'\lambda'}\;\delta_{\mu''\lambda''}.
 \label{unitary}\eeq
In order to avoid any possible ambiguity, Greek indices (whose values
are also 0 and 1) are used to label spinor components {\it after\/} the
unitary transformations. Note that the indices without primes refer to
the two particles of the first Werner pair (the only ones that are not
tested for spin up) and the primed indices refer to all the other
particles (that are tested for spin up).  The $V_{\nu\nu'\nu'',nn'n''}$
matrix elements of Bob's unitary transformation satisfy a relationship
similar to (\theequation). The generalization to a larger number of
Werner pairs is obvious.

After the execution of the unitary transformation (\ref{newrho}), Alice
and Bob have to test that all the particles, except those labelled by
the first (unprimed) indices, have their spin up. They discard any set
of $n$ Werner pairs where that test fails, even once. The density matrix
for the remaining ``successful'' cases is thus obtained by retaining, on
the right hand side of Eq.~(\ref{newrho}), only the terms whose primed
components are zeros, and then renormalizing the resulting matrix to
unit trace. This means that only two of the $2^n$ rows of the $U$
matrix, namely those with indices 000\ldots\ and 100\ldots, are relevant
(and likewise for the $V$ matrix).  The elimination of all the other
rows greatly simplifies the problem of optimizing these matrices. We
shall thus write, for brevity,

\beq U_{\mu 00,mm'm''}\to U_{\mu,mm'm''}, \eeq
where $\mu=0,1$. Then, on the left hand side of Eq.~(\ref{unitary}), we
effectively have two unknown vectors, $U_0$ and $U_1$, each one with
$2^n$ components (labelled by Latin indices  $mm'm''$). These vectors
have unit norm and are mutually orthogonal. Likewise, Bob has two
vectors, $V_0$ and $V_1$.  The problem is to optimize these four vectors
so as to make the expectation value of the Bell operator~\cite{BMR},

\beq C:=AB+AB'+A'B-A'B', \eeq
as large as possible.

The optimization proceeds as follows. The new density matrix, for the
pairs of spin-\mbox{$1\over2$} particles not yet tested by Alice and Bob
(that is, for the first pair in each set of $n$ pairs), is

\beq (\rho_{\rm new})_{\mu\nu,\sigma\tau}=N\,
 U_{\mu,mm'm''}\,V_{\nu,nn'n''}\;\rho_{mn,st}\;\rho_{m'n',s't'}\;
 \rho_{m''n'',s''t''}\,U^*_{\sigma,ss's''}\,V^*_{\tau,tt't''},\eeq
where $N$ is a normalization constant, needed to obtain unit trace
($N^{-1}$  is the probability that all the ``spin up'' tests were
successful). We then have~\cite{H3}, for fixed $\rho_{\rm new}$ and all
possible choices of $C$,

\beq \max\,[{\rm Tr}\,(C\rho_{\rm new})]=2\sqrt{M}, \label{M}\eeq
where $M$ is the sum of the two largest eigenvalues of the real
symmetric matrix $T^\dagger T$, defined by

\beq T_{pq}:={\rm Tr}\,[(\sigma_p\otimes\sigma_q)\,\rho_{\rm new}].
 \label{T} \eeq
Note that the matrix $T_{pq}$ is real, because the Pauli matrices
$\sigma_p$ and $\sigma_q$ are Hermitian, but in general it is not
symmetric (explicit formulas are given in the Appendix). Our problem is
to find the vectors $U_\mu$ and $V_\nu$ that maximize $M$.

At this point, some additional simplifying assumptions are helpful.
Since all matrix elements $\rho_{mn,st}$ are real, we shall restrict our
search to vectors $U_\mu$ and $V_\nu$ that only have real components.
(It is unlikely that higher values of $M$ can be attained by using
complex vectors, but this possibility cannot be totally ruled out
without a formal proof.)

Furthermore, the situations seen by Alice and Bob are completely
symmetric, except for the presence of opposite signs in the standard
expression for the singlet state:

\beq \textstyle{\psi=
 \left[{1\choose0}{0\choose1}-{0\choose1}{1\choose0}\right]
   \;/\sqrt{2}.} \eeq
The opposite signs can be made to become the same by redefining the
basis, for example by representing the ``down'' state of Bob's particle
by the symbol ${0\choose-1}$, {\it without\/} changing the basis used
for Alice's particle. This partial change of basis is equivalent a
substitution

\beq V_{\nu,nn'n''}\to(-1)^{\nu+n+n'+n''}\,V_{\nu,nn'n''},\eeq
on Bob's side. The minus signs in Eq.~(\ref{singlet}) also disappear,
and there is then complete symmetry for the two observers. It is
therefore plausible that, with that new basis, we have $U_\nu=V_\nu$.
Therefore, when we return to the original basis and notations, the
optimal $U_\nu$ and $V_\nu$ satisfy

\beq V_{\nu,nn'n''}=(-1)^{\nu+n+n'+n''}\,U_{\nu,nn'n''}.\eeq
We shall henceforth restrict our search to pairs of vectors that satisfy
this relation. (Without imposing this restriction, I checked, for a few
values of $x$, that the optimal vectors $U$ and $V$ indeed had that
symmetry property, when $n=2$ or 3. However, an exhaustive search for
$n=4$ would have exceeded the capacity of my workstation.)

After all the above simplifications, the problem that has to be solved
is the following: find two mutually orthogonal unit vectors, $U_0$ and
$U_1$, each with $2^n$ real components, that maximize the value of
$M(U)$ defined by Eqs.~(\ref{M}) and~(\ref{T}). This is a standard
optimization problem, which can be solved numerically, for example by
using the Powell algorithm~\cite{recipes}. Some care must however be
exercised.  The ortho\-normality constraints must be imposed in a way
that does not impede the convergence of the iterations. Moreover, there
is a continuous infinity of equivalent solutions, because the entire
experimental protocol, and therefore all the physical data, are
invariant under rotations around the quantization axes (namely, the axes
along which the ``spin up'' tests are performed). This means that a
substitution

\beq\begin{array}{lll}
 U_{0,mm'm''}&\to&U_{0,mm'm''}\,\cos\alpha-U_{1,mm'm''}\,\sin\alpha,\\
 U_{1,mm'm''}&\to&U_{0,mm'm''}\,\sin\alpha+U_{1,mm'm''}\,\cos\alpha,
\end{array}\eeq
for any real $\alpha$, does not affect the value of $M(U)$. Similar
transformations, with arbitrary angles, can also be performed on each
one of the $n$ other indices. Therefore the location of a maximum of
$M(U)$ is not a {\it point\/} in the $(2^{n+1}-3)$-dimensional parameter
space, but can lie anywhere on a $(n+1)$-dimensional {\it manifold\/}.

Since the function $M(U)$ is bounded, it must have at least one maximum.
It may, however, have more than one: there may be several distinct
$(n+1)$-dimensional manifolds on which $M(U)$ is locally maximal, each
one with a different value of the maximum. A numerical search by the
Powell algorithm~\cite{recipes} ends at one of these maxima, but not
necessarily at the largest one. The outcome may depend on the initial
point of the search. It is therefore  imperative to start from numerous
randomly chosen points in order to ascertain, with reasonable
confidence, that the largest maximum has indeed been found. (A curious
difficulty arises from the fact that Alice and Bob can always obtain
$\langle{C}\rangle=2$, irrespective of the quantum state, simply by
measuring the unit operator, so that all their results are $+1$. It is
important that the computer program used for optimization be immune to
such artifices.)\bigskip

\begin{center}{\bf III. RESULTS AND CONCLUSIONS}\end{center}

In all the cases that were examined, it was found that $M(U)$ has one of
its maxima for the following simple choice:

\beq U_{0,00\ldots}=U_{1,11\ldots}=1, \label{xor}\eeq
and all the other components of $U_0$ and $U_1$ vanish. Recall that the
``vectors'' $U_0$ and $U_1$ actually are {\it rows\/} $U_{000\ldots}$
and $U_{100\ldots}$ of the $2^n$-dimensional unitary matrix $U$ (the
other rows are irrelevant because of the elimination of all the
experiments in which a particle failed the spin-up test). In the case
$n=2$, one of the unitary matrices having the property (\ref{xor}) is a
simple permutation matrix that can be implemented by a ``controlled-{\sc
not}'' quantum gate~\cite{cnot}. The corresponding Boolean operation is
known as {\sc xor} (exclusive {\sc or}). For larger values of $n$,
matrices that satisfy Eq.~(\ref{xor}) will also be called {\sc
xor}-transformations.

It was found, by numerical calculations, that {\sc xor}-transformations
always are the optimal ones for $n=2$. They are also optimal for $n=3$
when the singlet fraction $x$ is less than 0.57, and for $n=4$ when
$x<0.52$. For larger values of $x$, more complicated forms of $U_0$ and
$U_1$ give better results. The existence of two different sets of maxima
may be seen in Fig.~1: there are discontinuities in the slopes of the
graphs for $n=3$ and~4, which occur at the values of $x$ where the
largest maximum of $\langle{C}\rangle$ passes from one of the manifolds
to the other one.

For $n=5$, a complete determination of $U_0$ and $U_1$ requires the
optimization of 64 parameters subject to 3 constraints, more than my
workstation could handle. I therefore considered only {\sc
xor}-transformations, which are likely to be optimal for $x\,
\mbox{\raisebox{-2pt}{{\scriptsize $\stackrel{\textstyle <}{\sim}$}}}\,
0.5$. In particular, for $x=0.5$ (the value that was used in Werner's
original work \cite{Wer}), the result is $\langle C\rangle=2.0087$, and
the CHSH inequality is violated. This violation occurs in spite of the
existence of an explicit LHV model that gives correct results if the
pairs are tested one by one. For $n\to\infty$, we expect the CHSH
inequality to be violated for $x>{1\over3}$ (that is, when the fidelity
is $F>{1\over2}$), because such pairs can be ``purified'' by the methods
of refs.~[15--18].

In summary, it has been shown that, even if a well defined LHV
model~\cite{Wer} can correctly predict all the statistical properties of
some pairs of particles when the particles are tested separately by two
distant observers, a definite nonlocal behavior (namely, a violation of
the CHSH inequality) may arise if several pairs are tested
simultaneously, provided that the particles held by each observer are
allowed to {\it interact locally before they are tested\/}. This result
is yet another example of the fact that more information can sometimes
be extracted by simultaneously testing several identically prepared
quantum systems, than by testing each one of them separately~\cite{PW}.
Note that, for such a phenomenon to occur, it is always necessary that
the distant observers exchange {\it classical\/}
information~\cite{Shimony}.\medskip

\begin{center}{\bf ACKNOWLEDGMENTS}\end{center}

I am grateful to A. Garg, N. D. Mermin and S. Popescu for helpful
comments, and to the authors of Refs.~[15--18] and \cite{Gisin96} for
advance copies of their articles. This work was supported by the Gerard
Swope Fund, and the Fund for Encouragement of Research.\medskip

\begin{center}{\bf APPENDIX}\end{center}

This Appendix explicitly lists all the components of the $T_{pq}$
matrix~(\ref{T}), when the density matrix $\rho_{mn,st}$ is real and
symmetric:

\beq T_{xx}=\rho_{00,11}+\rho_{01,10}+\rho_{10,01}+\rho_{11,00},\eeq
\beq T_{yy}=-\rho_{00,11}+\rho_{01,10}+\rho_{10,01}-\rho_{11,00},\eeq
\beq T_{zz}=\rho_{00,00}-\rho_{01,01}-\rho_{10,10}+\rho_{11,11},\eeq
\beq T_{xz}=\rho_{00,10}-\rho_{01,11}+\rho_{10,00}-\rho_{11,01},\eeq
\beq T_{zx}=\rho_{00,01}+\rho_{01,00}-\rho_{10,11}-\rho_{11,10}.\eeq
The other components vanish, because the Hermitian matrices 
$\sigma_p\otimes\sigma_q$ that have only one $y$-index are antisymmetric
(and pure imaginary). 

Recall that $M$ in Eq.~(\ref{M}) is the sum of the two largest
eigenvalues of $T^\dagger T$. One of the eigenvalues of this matrix
obviously is $T_{yy}^2$. The two others are obtained by diagonalizing
the symmetric matrix

\beq\left( \begin{array}{cc}
 T_{xx}^2+T_{zx}^2 & T_{xx}T_{xz}+T_{zx}T_{zz} \\
 T_{xz}T_{xx}+T_{zz}T_{zx} & T_{xz}^2+T_{zz}^2 \end{array}\right).\eeq
\vfill

\parindent 0mm
{\bf Caption of figure}\bigskip

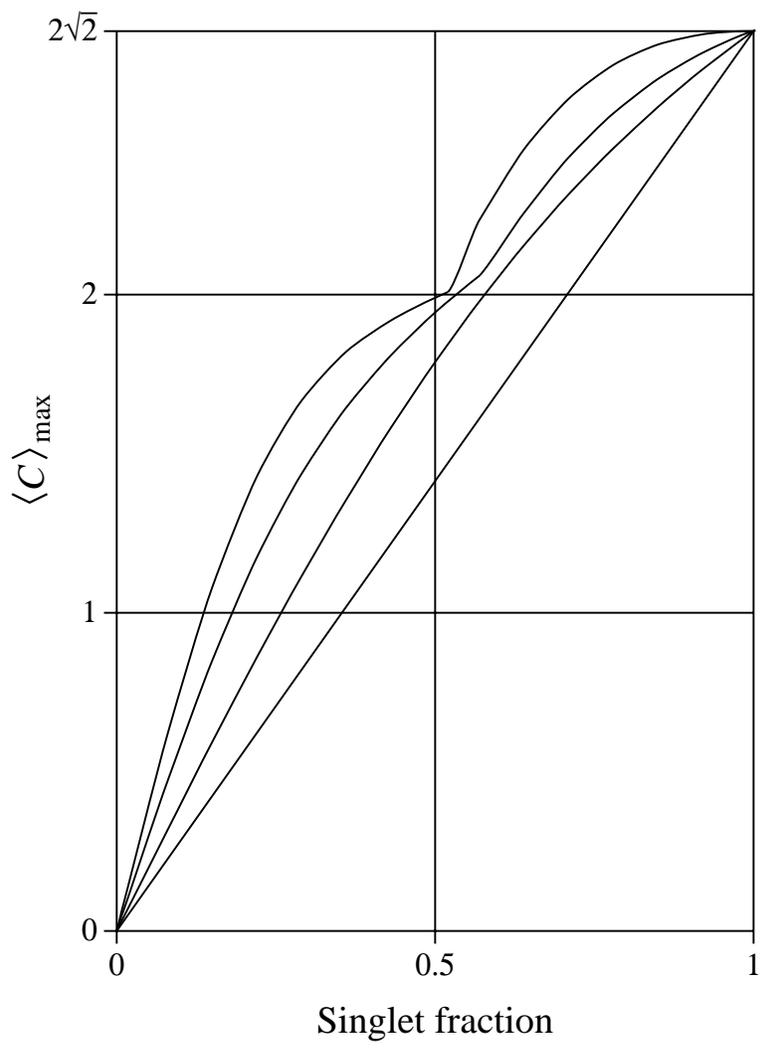
FIG. 1. \ Maximal expectation value of the Bell operator, versus the
singlet fraction in the Werner state, for collective tests performed on
several Werner pairs (from bottom to top of the figure, 1, 2, 3, and 4
pairs, respectively). The CHSH inequality is violated when
$\langle{C\rangle}>2$.

\end{document}